\documentclass{ws-mpla}
\usepackage[super]{cite}
\usepackage{graphicx}
\usepackage{color}
\begin{document}

\markboth{Partha Sarathi Debnath}
{Causal cosmology  with braneworld gravity including  Gauss Bonnet coupling }

%%%%%%%%%%%%%%%%%%%%% Publisher's Area please ignore %%%%%%%%%%%%%%
\catchline{}{}{}{}{}
%%%%%%%%%%%%%%%%%%%%%%%%%%%%%%%%%%%%%%%%%%%%%%%%%%%%%%%%%%%%%%%%%%%

\title{Causal cosmology  with braneworld gravity including  Gauss Bonnet coupling}

\author{ Partha Sarathi Debnath }

\address{Department of Physics,\\
 A. B. N. Seal College,\\
Coochbehar, Pin : 736 101, India.\\
parthasarathi6@hotmail.com}

\maketitle

\pub{Received (Day Month Year)}{Revised (Day Month Year)}

\begin{abstract}
Causal cosmological evolutions in  Randall Sundrum type II (RS) braneworld gravity with Gauss Bonnet  coupling and dissipative effects are discussed here. Causal theory of dissipative effects are illustrated by Full Israel  Stewart theory are implemented. We  consider the numerical solutions of evolutions  and analytic solutions as a special case for extremely non-linear field equation in Randall Sundrum type II braneworld gravity with Gauss Bonnet coupling.  Cosmological models admitting  Power law expansion, Exponential expansion and evolution in the vicinity of the stationary solution  of the  universe are investigated for  Full Israel  Stewart theory.  Stability of equilibrium or fixed points related to the dynamics  of evolution  in Full Israel Stewart theory in  Randall Sundrum type II  braneworld gravity together with   Gauss Bonnet coupling   are  disclosed here.

\end{abstract}

\ccode{PACS Nos.: 98.80.Jk;98.80.Es.}

\section{Introduction}	

Recent observational data suggest an accelerated expanding universe at the present epoch \cite{riess,Padma}. These cosmological observations also recommend  a specially flat universe with high accuracy. To cosmologist, it is a challenging theoretical problem to recognize the precise
 motivation for the present accelerated expansion. It is proposed that in the early universe there might be a phase of inflationary evolution. Theoretically inflation is understood considering scalar field of standard model. Inflaton are addressed by  Starobinsky \cite{staro,staro1} long ahead of beginning of inflation \cite{guth} is  recognized. However, in standard model the essential fields for present acceleration are  inaccessible. To address the present epoch the motivation of  dark energy including dark matter is established. According to PLANCK Collaboration\cite{Planck15} the contribution of dark energy is $\sim $ 69.4$\%$ and that of matter parts (mainly dark matter) are $\sim $ 30.6 $\%$ of total energy. At present,  it is one of the most challenging problem to address the dark universe in which dominating parts are dark matter and dark enrgry. 	The appraisal  of extensions  of general theory of relativity (GTR) are  considered to tackle the challenging issues  of the dark universe.  Some   proposals\cite{Capozz,Nojiri} are introduced by literature to  realize the accurate basics of the dark universe. 
	 Modified gravity such as   Gauss Bonnet \cite{Noji,Eliza} gravity, $f(T) $\cite{Cai,Bamba1} gravity , $f (R)$  \cite{Capo,Nojiri1} gravity, $f(R,T)$ \cite{harko,Tiwari,debnath1}  gravity and  Horava-Lifshitz \cite{Nishioka,Ran}  gravity are considered to realize the problem.
	Literature \cite{Witten} also consider another interesting and important modified theory of gravity   known as  braneworld gravity. 
    
In braneworld scenario \cite{hor1,hor2}, particles of standard model are restricted on brane surrounded by extra dimension bulk. In bulk matter and  gravity can only transmit. Randall Sundrum type II braneworld \cite{ran1,chen} gravity  of five dimension can simply  illustrate such circumstances. The Superstring/ M-theory  motivated  such Randall Sundrum model of braneworld gravity.  
 The  early stage of  universe described by Randall Sundrum typre II braneworld gravity can afford innovative type of evolution. In the brane theory, the five dimensional anti de Sitter space known as bulk surrounded practical universe  is considered as four dimensional brane. In a Randall-Sundrum type II scenario (RS II), spatially homogeneous and isotropic brane can be present in the extra dimensional anti de Sitter (AdS$_5$) bulk spacetime \cite{ran2}. At low energy the extra dimension support to detain graviton close to brane, so the general theory of relativity are recovered. However, at high energy extra dimension dominate as a result the graviton localization falls short that support a modification of Friedmann equations. The conformal field \cite{mal,wit}  theory also support this characteristics.  A conformal field theory \cite{haw, noj} coupled with normal gravity is counterpart of the Randall-Sundrum braneworld theory.  

 Einstein Hilbert (EH) action of five dimension support the Randall Sundrum braneworld gravity. The EH action  can attain quantum improvement 
 at high energies. To extend the braneworld theory  further in string theoretical background, it is essential to take account of  curvature invariant term ($R, \; R_{ab}R^{ab}, R_{abcd}R^{abcd}, ..$) in the bulk action \cite{noj2,noj3}. Within these modifications, the so called Gauss-Bonnet term can be included. The Gauss-Bonnet (GB $= R_{abcd}R^{abcd} - 4R_{ab}R^{ab} + R^2$, symbols usual significance) term has merely second derivatives of the metric of equation of motion which is   ghosts \cite{bar,cal} free and therefore important  \cite{capo2,capozzi} to consider. The most important quantum improvement in the heterotic string efficient action \cite{zwe,lov} shows Gauss Bonnet term.
 In EH bulk action \cite{mav,neu} with the presence of Gauss Bonnet term zero mode of gravitation localization on the brane is permitted. It is also found that the Gauss-Bonnet term  have a tendency to reduce the limitation of  Randall Sundrum braneworld theory. 
 Extensive studies of the Gauss-Bonnet brane world scenario \cite{cai1,deb} are considered to explain both the early inflation and the late acceleration \cite{capo3}.

Cosmological models with perfect fluid as a source of matters are considered in modified gravity theories. Modified theories of gravity without de Sitter solution or matter admitting inflation is unstable. Although perfect fluid satisfactorily describe matter sharing of the observed universe, however, the evolution of the universe in many phases guide to viscosity \cite{brevik2}. Viscosity may arises in different phases of evolution namely, radiation epoch, recombination epoch, superstrings in quantum epoch, graviton involved collision epoch, galaxies formation epoch \cite{mis, barr,pov,zim}. Therefore, it is necessary to incorporate viscosity in the evolution of the universe. Eckart \cite{eckart} theory is the first theory of viscosity, where concept of  non equilibrium thermodynamics is used in relativistic framework. However, Eckart \cite{eckart} theory suffers from causality and stability conditions. Israel and Stewart \cite{israel} extend a fully relativistic foundation  of the viscous theory considering second order deviation term to overcome the shortcomings.  In Israel and Stewart formalism phantom solution is also investigated in literature \cite{cruz}. In this paper, we provide an development  of the RS brane-world model incorporating  viscosity and Gauss Bonnet (GB) term in EH bulk action. The paper is planned as:  In section 2, relevant field equations in Randall Sundrum type II (RS) braneworld theory are set up with Gauss Bonnet (GB) term and viscosity. In section 3, we obtain  cosmological evolution in RS brane gravity with Gauss Bonnet (GB) term and causal viscous theory. Stability analysis  of the causal solution is also considered here. Finally, in sec. 4, we summarize  the results.

\section{ Relevant field equation for braneworld gravity }
The  5D Einstein Hilbert (EH) bulk action with GB term and 4D brane yields 
\begin{equation}
 S = \frac{1}{2 k_5^2}\int d^5 x \sqrt{- g_5} \left[- 2 \Lambda_5 +  R + \alpha ( R^2 - 4 R_{ab}  R^{ab}+R_{abcd}  R^{abcd})\right] - \int_{brane} d^4 x \sqrt{- g} \;\sigma 
\end{equation}
where $x^a = (x^\mu,z)$, $z$ is the co-ordinate for 5D,  $g_{ab} = g_{(5)ab} - n_a n_b$ is  metric of induced,  $n^{\alpha}$ is  unit normal on brane, $\sigma $ is positive tension on brane, $\alpha $ is the Gauss Bonnet coupling parameter with length$^2$ dimension   and cosmological constant in bulk is $\Lambda_5 (< 0)$. The 5D fundamental scale of energy is  $M_5$, with $k_5^2 = \frac{8 \pi}{M_5^3}$. The effective scale of energy which illustrate  gravity on brane for low energy is Planck scale $M_4 \sim 10^{16}$ TeV 
and usually $M_4 >> M_5$. The Gauss Bonnet coupling possibly consideration of the lowest order stringy improvement in 5D EH  bulk action and coupling parameter $\alpha > 0$. Here we consider $\alpha |R^2| << |R| $, with the intention that $ \alpha << l^2 ,$ and bulk  scale (curvature) is $l$ with $\mid R \mid \sim l^{-2}$.  One can recover RS braneworld model for considering $\alpha = 0$. 
The Friedmann brane in AdS$_5$ bulk with the presence of  $Z_2$ symmetry indicating  modified field equation for  (flat) GB braneworld  scenario is \cite{char,tuji,maed}
\begin{equation}
k_5^2 (\rho + \sigma) = 2\sqrt{H^2 + \mu^2} [3 - 4\alpha \mu^2 + 8\alpha H^2] ,
\end{equation}
where energy density for  matter fields on brane is $\rho$, Hubble parameter is $H$ and energy scale (effective) related to $l$ is $\mu$ $(\equiv l^{-1})$. 
One can rewrite the above mentioned equation in effective form \cite{lids}
\begin{equation}
H^2 = \frac{1}{4\alpha} \left[(1 - 4\alpha \mu^2) \cosh \left(\frac{2\psi}{3} - 1\right)\right],
\end{equation}
\begin{equation}
k_5^2 (\rho + \sigma) = \left[\frac{2(1 - 4\alpha \mu^2)^3}{\alpha}\right]^{1/2} \sinh \psi ,
\end{equation}
here $\psi$ stands for  dimensionless parameter of energy density.
  Using Eqs. (4)-(5),  we can achieve feature of GB scale (energy), 
$ m_{\alpha} = \left[ \frac{2(1-4\alpha \mu^2)^3}{\alpha k_5^4}\right]^{\frac{1}{8}} , $
so at  high energy (enough) GB regime illustrate to $\rho\gg m_{\alpha }^4$  for sinh$\psi \gg 1$. As GB coupling is an improvement of RS brane action, so  a restriction is forced  on $m_{\alpha}$ wherever $m_{\alpha}^4$ is larger than  RS brane scale (energy) $\sigma$.
We can obtain two important regimes to revise the dynamics of RS braneworld universe particularly at early evolution for enlarging values of   $\psi$ in Eq.(4).  The equivalent equations of field are :

{$\bullet$ } For  GB dominated regime, 
\begin{equation}
\rho >> m_{\alpha}^4 \Rightarrow H^2 \approx \left[\frac{k_5^2}{16\alpha} \rho\right]^{2/3} 
\end{equation}
which gives, $\rho = \rho_0 H^3$, where $\rho_0 = \frac{16\alpha}{k_5^2}$.

{$\bullet$ } For  RS dominated regime,
\begin{equation}
m_{\alpha}^4 >> \rho >> \sigma \equiv m_{\sigma}^4 \Rightarrow H^2 \approx \frac{k_4^2}{6\sigma} \rho^2 
\end{equation}
it shows $\rho = \rho_0 H$, where $\rho_0 = \left(\frac{6\sigma}{k_4^2}\right)^{1/2}$. 

In early time, at GB dominated regime the universe shows  evolution rate $H \sim \rho^{\frac{1}{3}}$, later in Randall-Sundrum  regime the rate is $H \sim \rho$ and finally, at low enough energy in standard evolution law, the rate is $H \sim \rho^{\frac{1}{2}}$.  \\
The  equation of conservation for energy momentum tensor yields   
\begin{equation}
\dot\rho = -3 H (\rho + p +\Pi),
\end{equation}
where energy density is denoted by $\rho $, isotropic nature of pressure is given by $p$  and  bulk viscous type of pressure is represented by $\Pi \;(\leq 0) $. Hence total efficient pressure $(p_{eff})$ on  braneworld gravity  illustrates  as $p_{eff}=p+\Pi$.  
 We consider a causal equation to address $\Pi$.  Here   $\Pi$ obeys subsequent causal transport equation \cite{israel} of Full Israel Stewart theory 
\begin{equation}
\tau\dot{\Pi} + \Pi =-3\zeta H -\frac{\Pi\tau}{2}\left(3H+\frac{\dot{\tau}}{\tau}-\frac{\dot{\zeta}}{\zeta} -\frac{\dot{T}}{T} \right),            
\end{equation}
where bulk viscous coefficient is  $\zeta (> 0)$,  time of relaxation is  $\tau (\geq 0)$  and $T (>0)$ indicates the universe's temperature.   The time of relaxation  $(\tau)$  and  bulk viscous  coefficient  $(\zeta)$  are defined \cite{brevik,paul2,jou} respectively as   
\begin{equation}
\tau=\beta \rho ^{r-1},\;\;\;\;\;\zeta=\beta_1 \rho^s ,
\end{equation}
where $s$, $r$, $\beta$ and $\beta_1$    are  positive constants.
 With the choice of $\tau$ and $\zeta$ the viscous signal propagates with speed $v=\sqrt{\frac{\zeta}{\rho\tau}}=\sqrt{\frac{\beta_1}{\beta}\rho^{s-r}}$. Physically viable solutions are permitted $(v\leq 1)$ for (i) $\beta \geq \beta_1$ and $r\geq s $ at higher values of energy density $(\rho\geq 1)$, (ii) $\beta\geq \beta_1$ and $s\geq r $ at smaller values of energy density $(\rho\leq 1)$. An inflationary phase remain possible for $r=s$. The reasonable values of parameters such as  $r=s=1$ describe radiative fluid and $r=s=\frac{3}{2}$ corresponds to a string dominated universe \cite{jou}.
 The affirmative values of entropy generation are established due to positive signed of $\zeta$. The relation  between isotropic nature of  pressure  with energy density yields 
\begin{equation}
p\;=\omega \;\rho .   
\end{equation}
Here $\omega  $ is known as the   equation of state parameter. In this article we consider that $\omega$ is a constant parameter. Where the values of $\omega\leq 1$ represent causal solution, the values of $\omega>-1$ represent quintessence fluid,  the values of $\omega<-1$ represent phantom model and  $\omega=-1$ represents a  vacuum solution.  
Another vital cosmological parameter to study evolution of  universe is the parameter of deceleration $(q)$. 
The parameter $q$ is defined as 
\begin{equation}
 q =- \frac{\ddot{a}a}{\dot{a}^2} = - \frac{\dot{H}}{H^2}-\;1 ,
\end{equation}
where $a$ is the scale factor. The accelerated phases of the universe are characterized by $q<0$, the decelerated phases are characterized by $q>0$ and $q=0$ represent neither acceleration nor deceleration type evolution.   
\section{ Causal  solutions :}  
The following section illustrates  cosmological solutions of  Full Israel Stewart (FIS) theory in brane-world including Gauss-Bonnet term. 
	In Full Israel Stewart theory, transport equation yields 
\[
	\ddot{H} +\frac{\epsilon}{2}\left[r-s+\frac{1}{1+\omega}-\frac{4}{\epsilon}\right]\frac{\dot{H^2}}{H} +\frac{3(1+\omega)}{\epsilon \beta}\rho_{0}^{1-r}H^{2+\epsilon-\epsilon r} - \frac{9\eta}{\epsilon}\rho_{0}^{s-r}H^{\epsilon(s-r)+3} 
	\]
		\begin{equation}
		+\left[ 3 +\frac{1}{\beta} \rho_{0}^{1-r} H^{\epsilon-\epsilon r-1}+\frac{3}{2}(r-s)(1+\omega)\right] H\dot{H} +\frac{9(1+\omega)H^3}{2\epsilon}=0.
	\end{equation}
The  universe's temperature  is defined as $T =T_0 \rho^{\frac{\omega}{1+\omega}}$, where $T_0$ is constant  and constant parameter $\eta=\frac{\beta_1}{\beta}$. Parameter $\epsilon$ has two values either 3 or 1.  In GB dominated regime $\epsilon=3$ and for RS dominated regime $\epsilon=1$.
Field Eq. (12) represents extremely nonlinearity, so very hard to get a wide-ranging analytical solution of known form. 
To get exact analytical solutions in GB dominated and RS dominated regime as special case we regard as  $r=\frac{\epsilon-1}{\epsilon}$ for simplicity, Eq. (12)  yields
\begin{equation}
H\ddot{H}+ b_1 \dot{H}^2 + b_2 H^2\dot{H} + b_3 H^4 -b_4H^{5+(s-1)\epsilon}=0,
\end{equation}
where $b_1=\frac{\epsilon}{2}(1-s+\frac{1}{1+\omega}-\frac{5}{\epsilon})$, $b_2=3+\frac{1}{\beta}\rho_0^{\frac{1}{\epsilon}}+\frac{3}{2} (1+\omega)(1-s-\frac{1}{\epsilon})$, $b_3= \frac{9(1+\omega)}{2\epsilon} + \frac{3(1+\omega)}{\epsilon\beta}\rho_0^{\frac{1}{\epsilon}}$ and $b_4=\frac{9\eta}{\epsilon}\rho_0^{s+1-\frac{1}{\epsilon}}$. To obtain evolution of parameters $a$ and $H$ analytically from Eq. (13)  for RS  dominated regime and GB dominated regime,  we regard as subsequent particular cases \\
$\bullet$ Case (i) $s=\frac{\epsilon-1}{\epsilon}$:  Here we consider  $y$ and $\theta$ as a new set of  variables  defined as  $y^n=H$ and $\theta= 3 m \;t$ where $n=\frac{1}{b_1 +1}$ and $m=\frac{b_2}{3}$. Equation (13) yields 
\begin{equation}
\frac{d^2 y}{d\theta^2} + y^n \frac{dy}{d\theta} + \frac{\beta_2}{n+1} y^{2n+1}=0,
\end{equation}
 where $\beta_2=\frac{(b_3-b_4)(n+1)}{9m^2 n}$. By means of  local transformation of variables $z=\frac{1}{n+1} y^{n+1}, \; d\phi = y^n d\theta$, Eq. (14) yields a second order linear differential equation 
\begin{equation}
\frac{d^2 z}{d\phi^2} + \frac{dz}{d\phi} + \beta_2 z=0.
\end{equation}
Using the Eq. (15) the parametric nature of Hubble parameter ($H$)  yields
\begin{equation}
H=\left[ c_1 e^{p_+\phi} + c_2 e^{p_-\phi} \right]^{\frac{n}{n+1}}, \;\; \; t=\frac{1}{3m}\int \frac{d\phi}{ H(\phi)}.
\end{equation}
  Here $p_{\pm}=\frac{1}{2} \left[-1\pm \sqrt{1-4\beta_2} \right]$ and $c_1$, $c_2$ are constants and  we also make a note that $p_+ > 0$ for $\beta_2<0$. An exponential inflation in parametric nature of time  is permitted with positive singed  of $p_{\pm}$, $n,\; m$,  $c_1$, $c_2$.\\
	$\bullet$ Case (ii) $s=\frac{\epsilon-3}{\epsilon},\;\frac{\epsilon-2}{\epsilon},\;\frac{\epsilon-1}{\epsilon}$: In these particular cases from Eq. (13) one can obtain emergent universe  \cite{mukherjee,debnath} solution and  the scale factor  yields 
	\begin{equation}
a(t)= a_i \left[a_1 + e^{a_2 t }\right]^{\delta},
\end{equation}
	where $a_i$, $a_1$, $a_2$ and $\delta$ are constants with positive values. In case of $s=\frac{\epsilon-3}{\epsilon}$, emergent universe solution is permitted for $b_1=b_4 -1$, $b_2= \frac{2b_4}{\delta}+\frac{1}{\delta}$ and $b_3 = \frac{b_4}{\delta^2}$.  In case of $s=\frac{\epsilon-2}{\epsilon}$, emergent universe solution is permitted for $b_1=-1$, $b_2= b_4+\frac{1}{\delta}$ and $b_3 = \frac{b_4}{\delta}$. In case of $s=\frac{\epsilon-1}{\epsilon}$, emergent universe solution is permitted for $b_1=-1$, $b_2= \frac{1}{\delta}$ and $b_3 = b_4$.  We note that emergent universe solution is permitted in GB regime for $r=\frac{3}{2}$ and $s=\frac{3}{2},\; \frac{1}{3},\; 0$. However, in RS regime, emergent universe solution is permitted for $r=0$ and $s=0$. \\
	 It is worth to note that rearranging Eq. (12), the field equation for zero viscosity ($ \beta =\beta_1=0$)  in brane-world including Gauss-Bonnet term  leads $\dot{H}+\frac{3}{\epsilon}(1+\omega)H^2=0$, where $\epsilon=3$ for GB dominated regime and $\epsilon=1$ for RS dominated regime.  \\
	In the following subsections we study  cosmological solution in GB and RS regime separately.
	\subsection{GB dominated regime :}
	Field equation in GB  dominated regime for  the causal transport is obtained form  Eq. (12) by putting $\epsilon=3$, which  yields
	\[
	\ddot{H} +\frac{3}{2}\left[r-s-\frac{1+4\omega}{3(1+\omega)}\right]\frac{\dot{H^2}}{H} +\left[  \frac{3}{2}(1+\omega)(r-s)+3+\frac{\rho_0^{1-r}}{\beta} H^{2-3r}\right] H\dot{H} 
	\]
		\begin{equation}
	+H^3\left[\frac{1+\omega}{\beta} \rho_0^{1-r} H^{2-3r}- 3\eta\rho_0^{s-r}	H^{3(s-r)}+ \frac{3}{2}(1+\omega) \right]=0. 
	\end{equation}	
	Although it be very complicated to get analytical solution of wide-ranging form in GB dominated regime from Eq.(18), however, we  be able to get  solutions of important cosmological  parameters in numerical form, for example,  Hubble parameter $(H)$, scale factor $(a)$.
	We study numerically the evolution of the universe as follows:\\
	(i) The variation of $H$ with $t$ in GB dominated regime for different values of bulk viscous constant $(\beta)$ is plotted in the Fig. 1. 
	 It shows that the evolution of $H$ has declining nature characteristics with cosmic time.  It is found that at a specified moment the smaller values of $H$   lead to  smaller values of bulk viscous constant. \\
	\begin{figure}[ph]
\centerline{\includegraphics[width=2.0in]{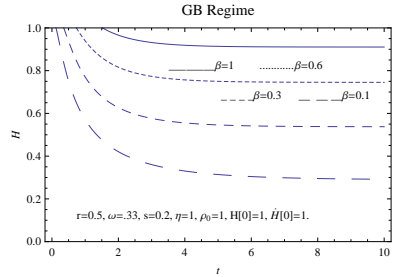}}
\vspace*{8pt}
\caption{Illustrates $H$  vs.  $t$ in GB dominated regime for  various $\beta$ with other known  parameters. \protect\label{fig1}}
\end{figure}
(ii) The variation of $a$ with $t$ in GB dominated regime  for different values of bulk viscous constant $(\beta)$ is plotted in the Fig. 2. It is evident that $a$ has increasing nature characteristics  with the cosmic time. It is also found that at a specified moment  the higher  values of scale factor lead to  higher values of bulk viscous constant.
\begin{figure}[ph]
\centerline{\includegraphics[width=2.0in]{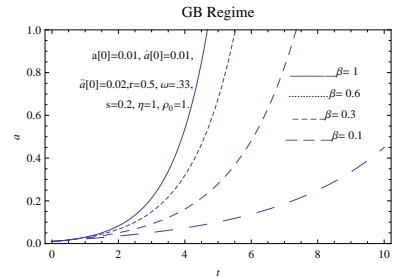}}
\vspace*{8pt}
\caption{Illustrates   $a$  vs.  $t$ in GB dominated regime for  various $\beta$ with other known  parameters. \protect\label{fig2}}
\end{figure}
\\To study analytic solutions of known form for causal cosmology in GB regime we consider following special cases.  
\subsubsection{Power-law expansion:} In power-law  model,  scale factor of the universe evolves as $a(t)=a_0 t^D$, where $a_0$ and exponent $D$ indicate constant parameters. In this model,  deceleration parameter ($q$) yields $q=\frac{1-D}{D}$, it shows accelerated expansion for $D>1$.	In GB regime, the energy density and the viscous stress can be  represented    as respectively 
	\begin{equation}
	\rho=\rho_{0} D^3 t^{-3}, \;\;\;\; \;\;\Pi = -\rho_{0} D^2((1+\omega)D-1) t^{-3} ,
	\end{equation}
	where exponent $D>\frac{1}{(1+\omega)}$ indicate physically relevant solutions ($\Pi<0$). 
 In GB regime for power-law expansion with Full Israel Stewart (FIS)  theory the field Eq. (18) yields
   \begin{equation}
	A_1 + A_2 t^{3r-2} + A_3 t^{3(r-s)} =0,
	\end{equation}             
where $A_1=\frac{3}{2}(r-s+\frac{1}{1+\omega})-\frac{3D}{2}((1+\omega)(r-s)+2)$, $A_2=\frac{1}{\beta}\rho_{0}^{1-r} D^{3-3r} (D(1+\omega)-1) $ and $A_3=-3\eta\rho_{0}^{s-r} D^{3s-3r+2}$. In GB regime power-law solution is permitted for the subsequent cases:\\
Case (i) $ r\neq \frac{2}{3}$ and $s\neq r$: Power-law solution is permitted for $A_1=A_2=A_3=0$,  which yields $\rho_{0}=0$. It shows a physically non realistic power-law solution.\\
Case (ii) $ r\neq \frac{2}{3}$ and $s=r$: Power-law solution is permitted for $A_1+A_3=0$ and $A_2=0$. It  gives the power-law exponent $D=\frac{1}{1+\omega}$. It indicates that  bulk viscous stress $(\Pi)$ becomes zero, which is not physically acceptable solution.\\
Case (iii) $ r\neq \frac{2}{3}$ and $s=\frac{2}{3}$: In this case, power-law solution is permitted for $A_1=0$ and $A_2+A_3=0$. It  yields the bulk viscus constant $\beta_1= \frac{(D(1+\omega)-1)\rho_{0}^{\frac{1}{3}}}{3D}$ and the power-law exponent  $D=r+\frac{1-2\omega}{3(1+\omega)}$ or $\frac{1}{1+\omega}$. We note that physically viable solution is permitted for $D=r+\frac{1-2\omega}{3(1+\omega)}$. In this type of dissipative process the values of power law exponent $(D)$ depend on both bulk viscous parameter $r$ and  parameter of state $(\omega)$. However for $\omega=\frac{1}{2}$, exponent $D$ depends on $r$ particularly. 
 If we choose $r=\frac{5}{3}+\frac{\omega}{1+\omega}$ as a special case, it gives $D=2$. In this special case, the  deceleration parameter becomes  $q=-0.5$ which leads to current observed values of  $q$  \cite{sahoo}. \\
Case (iv) $r=s=\frac{2}{3}$:  In this case,  power-law evolutions  are permitted   in GB dominated  regime for $A_1+A_2+A_3=0$ which leads to
\begin{equation}
\left[1+\omega+\frac{2(1+\omega)\rho_{0}^{\frac{1}{3}}}{3\beta}-2\eta \right]D^2+ \left[2+\frac{2\rho_{0}^{\frac{1}{3}}}{3\beta}\right]D+\frac{1}{1+\omega}=0.
\end{equation}
 Power law type inflation  is permitted ($D>>1$) in Full Israel Stewart theory for  known values of other parameters used in Eq. (21). Figure 3 illustrates the variation of $D$ vs. $\omega$ for various  $\beta$. The shadow region of the Fig. 3 represents $\Pi>0$, which is unsuitable for power law expansion. The figure demonstrates that the power law  inflation in GB dominated regime is motivated for higher $\omega$ and  higher $\beta$.
\begin{figure}[ph]
\centerline{\includegraphics[width=2.0in]{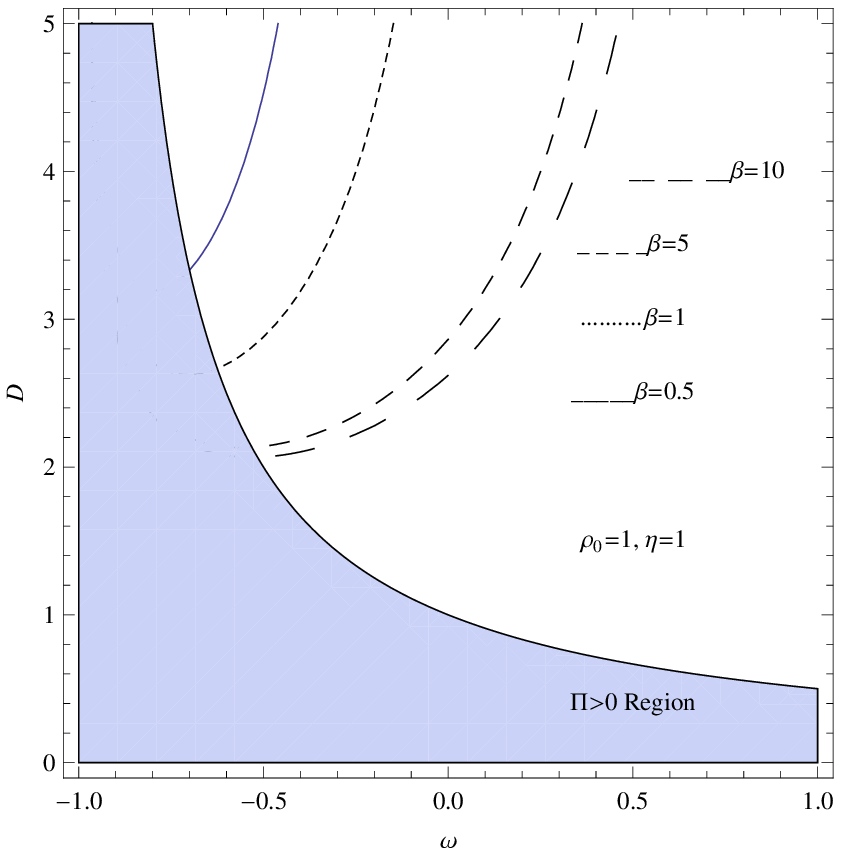}}
\vspace*{8pt}
\caption{Illustrates   $D$  vs  $\omega$  in GB regime for  various $\beta$ with other known  parameters. \protect\label{fig3}}
\end{figure}
\\
	 In the absence of bulk viscosity ($ \beta =\beta_1=0$) the field equation ($\dot{H}+(1+\omega)H^2=0$) in GB regime 
		yields power law type expansion ($a(t)=a_0 t^D$) unless $\omega=-1$, where power law exponent $D=\frac{1}{(1+\omega)}$. In GB regime without bulk viscosity, power law type accelerated expansion ($D>1$) is permitted for $0>\omega>-1$. 
\subsubsection{Exponential model :} 
Exponential type cosmic evolution  $(a(t)\sim e^{H_1 t})$ is  permitted in RS brane with GB term for $\ddot{H}=\dot{H}=0$ or $H=const.=H_1$ in Eq. (18). Exponential expansion leads de Sitter type expansion for $H_1>0$ even in the presence of matters.
For exponential expansion type model the Hubble's parameter in GB regime  yields
\begin{equation}
\frac{(1+\omega)}{\beta}\rho_{0}^{1-r}H_1^{2-3r} - 3\eta\rho_{0}^{s-r}H_1^{3(s-r)} +\frac{3(1+\omega)}{2}=0.
\end{equation}
Several possibility arises to implement an exponential expansion in  GB regime.   Subsequent cases are considered  for simplicity:\\
Case (i)$s\neq \frac{2}{3}$  and $r=s$ : In this particular case exponent $(H_1)$ of exponential expansion  reduces to $H_1 =\left[\frac{3\beta(2\eta-1-\omega)}{2(1+\omega)\rho_0^{1-r}} \right]^{\frac{1}{2-3s}}$. The de Sitter type expansion is permitted for  $2\eta-1>\omega>-1$. The causality condition ($\omega\leq1$) of the solution implies  following  constraint on the upper range of parameter $\eta$, which is $\eta\leq1$.\\
Case (ii) $r\neq\frac{2}{3}$ and $s= \frac{2}{3}$ : The exponent $(H_1)$ of exponential expansion  yields $H_1 =\left[\frac{3(1+\omega)\rho_0^{r-\frac{2}{3}}}{2(3\eta-\frac{1+\omega}{\beta}\rho_0^{\frac{1}{3}})} \right]^{\frac{1}{2-3r}}$. Exponential expansion is permitted for  (i) $\omega>-1$ and $\beta_1 > \frac{(1+\omega)\rho_0^{\frac{1}{3}}}{3\eta}$,  (ii) $\omega<-1$ and $\beta_1 < \frac{(1+\omega)\rho_0^{\frac{1}{3}}}{3\eta}$. However, former one represents exponential solution for quintessence like fluid and the later one represents that for phantom like fluid.  \\
Case (iii) $r=\frac{2}{3}$ and $s\neq \frac{2}{3}$ : In this particular case exponent $(H_1)$ of exponential expansion  reduces to $H_1 =\left[\frac{(1+\omega)\rho_0^{\frac{2}{3}}}{3\eta\rho_0^{s}}\left(\frac{\rho_0^{\frac{1}{3}}}{\beta}+\frac{3}{2} \right) \right]^{\frac{1}{3s-2}}$. 
Here the de Sitter type expansion is permitted  for  $\omega>-1$, $\beta>0$ and $\eta>0$. \\
 In the absence of bulk viscosity ($ \beta =\beta_1=0$) the field equation ($\dot{H}+(1+\omega)H^2=0$) in GB regime 
		yield exponential expansion ($a(t)=e^{H_1 t}$) for $\omega=-1$, which permits accelerated expansion for $H_1>0$.\\
 We now study stability of the cosmic evolution  in the  GB dominated regime with causal  dissipation.  The exponential evolution admits  accelerating phase for $H_1>0$. 
In the causal cosmology the cosmological evolution is directed by differential equation which is second order in nature. To revise stability of cosmic evolution due to equilibrium points or fixed points in GB dominated regime, one can rewrite Eq. (18) in term of two autonomous first order differential equations which are 
\[ \dot{H}=y ,\]
\[ \dot{y}= P(y,H)= -\frac{3}{2}\left[r-s-\frac{1+4\omega}{3(1+\omega)}\right] \frac{y^2}{H}-\left[ \frac{3}{2}(1+\omega)(r-s)+3+\frac{\rho_0^{1-r}}{\beta}H^{2-3r} \right] H y    \]
\begin{equation}
-\left[\frac{1+\omega}{\beta}\rho_0^{1-r} H^{5-3r}-3\eta \rho_0^{s-r} H^{3(s-r+1)} +\frac{3}{2}(1+\omega)H^3  \right]  .
\end{equation}                
The phase point $P(0,H_1)=0$ or the equilibrium point is described by  $\dot{y}=y=0$ in Eq. (23). 
Considering the expansion of Taylor's and  linear approximation \cite{smith} about the fixed points, the Eq. (23) becomes
\begin{equation}
\dot{y}= e \;H+ h\; y \;.
\end{equation} 
Where the constants are given by  $e=-(1+\omega)H_1^2\left[\frac{2-3s}{\beta}\rho_0^{1-r} H_1^{2-3s} + \frac{9}{2}(r-s) \right]$ and $h=-H_1 \left[\frac{3}{2}(1+\omega)(r-s)+ 3+ \frac{1}{\beta} \rho_0^{1-r} H_1^{2-3r} \right]$. The  equilibrium points associate to the de Sitter type  evolution  in GB dominated regime are characterized in Table 1.  The  exponential expansions corresponding stable acceleration are realized by (i) $r\geq s$, $s< \frac{2}{3}$, $2\eta-1\geq \omega > -1$ and  $0<\eta\leq 1$ (ii)  $r > s$, $s\leq \frac{2}{3}$, $2\eta-1\geq \omega > -1$ and  $0<\eta\leq 1$. 
\begin{table}[ht]
\tbl{ Stability  of fixed points associated to de Sitter type expansion in GB dominated regime for causal solution. }
{\begin{tabular}{@{}cc@{}} \toprule
$Constraints\; of\; parameters$ & $Type \;of\;equilibrium \;points $   \\  \colrule
$r\geq s$, $s<\frac{2}{3}$, $2\eta-1\geq \omega > -1$, and  $0<\eta\leq 1$  & $stable\; attractor$ \\
$r > s$, $s\leq \frac{2}{3}$, $2\eta-1\geq \omega > -1$ and  $0<\eta\leq 1$ & $stable\; attractor$ \\
$r < s$, $s\geq \frac{2}{3}$, $ \omega > -1$ and  $H_1 > 0$  & $unstable\; saddle$ \\
$r=s$, $s=\frac{2}{3}$ and  $H_1 > 0$   & $centre$ \\
$r<s$, $s<\frac{2}{3}$, $\omega>-1$, $H_1>\left[\frac{9(s-r) \beta}{2\rho_0^{1-r}}\right]^{\frac{1}{2-3s}} $& $stable\; attractor$ \\\botrule
\end{tabular}}
\end{table}
\subsubsection{Evolution in the vicinity of the stationary solution : }
In Eq. (22)  $H=H_1$ implies exponential inflationary expansion in GB regime with a constant rate given by $H_1$. We want to examine analytically in detail the behaviour of both Hubble parameter  ($H$) and  scale factor $(a(t))$ when cosmological evolution is close to any stationary solution $(H_1)$. The behaviour of  the Hubble parameter in  vicinity of  stationary solutions are studied by setting $H=H_1 +\chi$ and $\chi<<H_1$. Using the relation the behaviour of scale factor in the vicinity of stationary solution yields $a(t)=a_0' e^{H_1t+\int\chi dt}$, where $a_0'$ is a constant parameter. By Setting $H=H_1+\chi$, with $|\chi<<H_1|$ and after linearization Eq.(18)  yields,
\begin{equation}
\ddot{\chi}+ \lambda_1\dot{\chi}+  \lambda_2 \chi =0,
\end{equation}
 where we consider  the constant parameters $\lambda_1=(\frac{7+4\omega}{2}+ \frac{H_1^{2-3r}}{\beta\rho_{0}^{r-1}})H_1$, $\lambda_2= (\frac{(1+\omega)(2-3s) }{\beta \rho_{0}^{r-1}}  H_1^{2-3r}+\frac{3(1+4\omega)}{2}) H_1^2$ and $r-s-\frac{1+4\omega}{3(1+\omega)}=0$. The solution of above Eq. (25) yields 
\begin{equation}
\chi(t)=\chi_1 e^{\delta_+ t} +\chi_2 e^{\delta_-t},
\end{equation}
where $\chi_1$ and $\chi_2$ are constants which depend on initial conditions. Here $\delta_+$ and $\delta_-$ are the roots satisfying following relation
\begin{equation}
\delta_{\pm}=\frac{\lambda_1}{2}\left[-1\pm\sqrt{1-\frac{4\lambda_2}{\lambda_1^2}}\right].
\end{equation}
Several possibilities arise and  following cases are consider for simplicity:\\  
Case (i) weak damping $(\lambda_1 ^2<4\lambda_2)$ : The quantity below  square root in Eq. (27) turns into negative and the corresponding solution for $\chi$ yields  $\chi(t)=\chi(0)e^{-\frac{\lambda_1t}{2}} cos(\frac{1}{2}\sqrt{\lambda_1^2-4\lambda_2}\;t)$, where $\chi(0)$ is an initial value constant. Hence Hubble parameter ($H$) exhibits an oscillatory damped behaviour of frequency $n=\frac{1}{2}\sqrt{\lambda_1^2-4\lambda_2}$ around the stationary solution $(H=H_1)$. The damped oscillatory solution yields inflation  with oscillation given by the term $\int \chi dt={e^{-\frac{\lambda_1t}{2}}[a_1 cos(\frac{1}{2}\sqrt{\lambda_1^2-4\lambda_2}\;t+ a_2 sin(\frac{1}{2}\sqrt{\lambda_1^2-4\lambda_2}\;t]}$, where $a_1,\; a_2$ are constants.  This nearly stationary solution has the curious feature \cite{pavon}.   In GB regime for $r=\frac{2}{3}$, one can obtain a damped oscillatory behaviour around $H_1$ for the constraint $\frac{\rho_0^{\frac{1}{3}}}{\beta}<\frac{12\omega-3}{2} \pm 2 \sqrt{8\omega^2-2\omega-1}$. To show stable evolution in the vicinity of the stationary solution, one can plot $H$ versus $t$ for  $|\chi<<H_1|$. Here the  solution will be a stable spiral for  $s\leq\frac{2}{3}$ and $\omega>-\frac{1}{4}$. Figure 4 shows $H$ vs. $t$ for a specified other parameters in GB regime. The figure shows stable expansion of the universe with time around a equilibrium point $(H_1)$ for different values of $\chi$ ( $<< H_1$), $\lambda_1^2<4\lambda_2$ and $\lambda_1>0$.
\begin{figure}[ph]
\centerline{\includegraphics[width=2.0in]{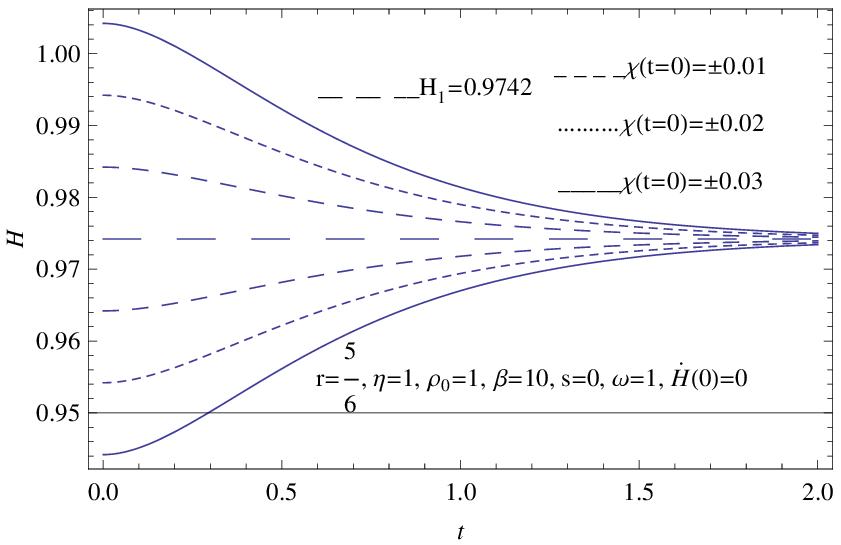}}
\vspace*{8pt}
\caption{Illustrates  $H$  vs  $t$ in GB dominated regime for various  $\chi(t=0)$ with a known set of  parameters. \protect\label{fig4}}
\end{figure}\\  
Case (ii) strong damping $(\lambda_1 ^2 > 4\lambda_2)$ : The quantity below  square root in Eq. (27) is real and both roots in Eq. (27) shall be real. Furthermore,  the quantity within the square bracket in Eq. (27) is negative for $\lambda_2>0$. Hence both  solutions  shall be a stable node for  $s\leq\frac{2}{3}$ and $\omega>-\frac{1}{4}$.  The   solution yields inflation  with  the term $\int \chi dt=a_1'e^{\delta_+ t} + a_2'e^{\delta_- t} $, where $a_1'$ and $a_2'$ are constants. \\
Case (iii) critical damping  $(\lambda_1 ^2 = 4\lambda_2)$ : In this case $\delta_+ =\delta_-=-\frac{\lambda_1}{2}$, the solutions are given by $\chi(t) = \left( a_1^{''} + a_2^{''} t \right) e^{-\frac{\lambda_1 t}{2}}$, where $a_1^{''}$ and $a_2^{''}$ are  constants. The solutions resemble those for strong damping and the solutions show a  stable node  for  $s\leq\frac{2}{3}$ and $\omega>-\frac{1}{4}$. \\
Case (iv) $\lambda_2<0$: In this case $\delta_+$ and  $ \delta_-$ are real but opposite sign. The solutions show a unstable saddle for (i) $s \geq \frac{2}{3}$ and $-\frac{1}{4} > \omega > -1$, (ii) $s\leq \frac{2}{3}$ and $\omega <-1$. \\
 In the absence of bulk viscosity ($ \beta =\beta_1=0$) the field equation in GB regime yields  $\dot{H}+(1+\omega)H^2=0$ which is a first order differential equation of $H$. In the absence of bulk viscosity damped oscillatory behaviour of Hubble parameter ($H$)  in the vicinity of stationary solution ($H_1$) is not permitted due to the order of the field equation.
\subsection{RS dominated regime:}
Field equation in  braneworld gravity (RS II) with FIS theory is obtained from Eq. (12) by setting $\epsilon =1$, which yields 	
\[
	\ddot{H} +\frac{1}{2}\left[r-s-\frac{3+4\omega}{(1+\omega)}\right]\frac{\dot{H^2}}{H} +\left[  \frac{3}{2}(1+\omega)(r-s)+3+\frac{\rho_0^{1-r}}{\beta} H^{-r}\right] H\dot{H} 
	\]
		\begin{equation}
	+3H^3\left[\frac{(1+\omega)}{\beta} \rho_0^{1-r} H^{-r}- 3\eta\rho_0^{s-r}	H^{(s-r)}+ \frac{3}{2}(1+\omega) \right]=0.
	\end{equation}
	The field Eq. (28) is very non linear to acquire a wide-ranging analytical solution in RS regime. Though, we can study relevant numerical solutions of cosmological  parameters for instance Hubble parameter $(H)$, scale factor $(a)$.	
	We study numerically the evolution of the universe as follows:\\
	(i) The variation of $H$ with $t$ in RS dominated regime for different values of bulk viscous constant $(\beta)$ is plotted in the Fig. 5. 
	 It shows that the evolution of $H$ has declining nature characteristics with cosmic time.  It is found that at a specified moment the smaller values of $H$   lead to  smaller values of bulk viscous constant. \\ 
\begin{figure}[ph]
\centerline{\includegraphics[width=2.0in]{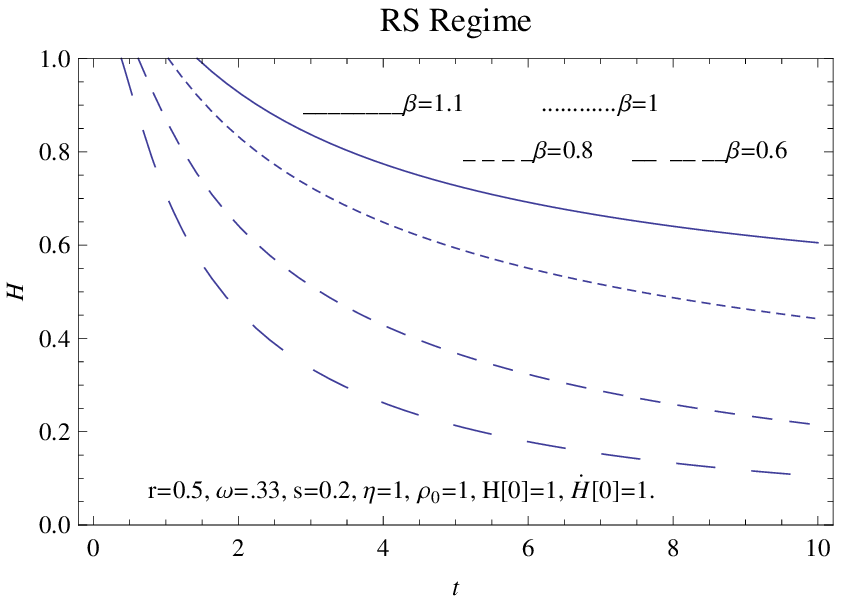}}
\vspace*{8pt}
\caption{Illustrates $H$  vs.  $t$ in RS dominated regime for  various $\beta$ with other known  parameters. \protect\label{fig5}}
\end{figure}
	(ii) The variation of $a$ with $t$ in RS dominated regime  for different values of bulk viscous constant $(\beta)$ is plotted in the Fig. 6. It is evident that $a$ has increasing nature characteristics  with the cosmic time. It is also found that at a specified moment  the higher  values of scale factor lead to  higher values of bulk viscous constant.  
\begin{figure}[ph]
\centerline{\includegraphics[width=2.0in]{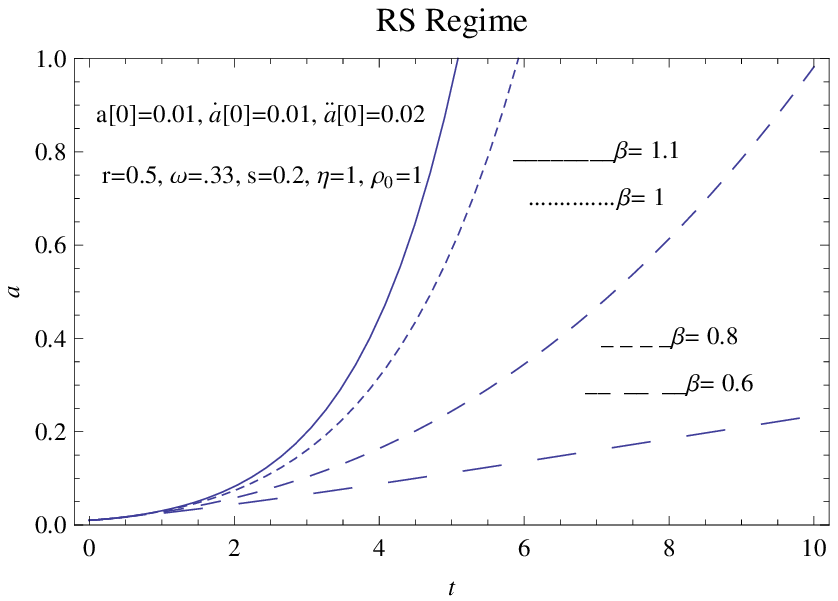}}
\vspace*{8pt}
\caption{Illustrates $a$  vs.  $t$ in RS dominated regime for  various $\beta$ with other known  parameters. \protect\label{fig6}}
\end{figure}
\\To study analytic solution of known form for causal cosmology in RS regime  we consider following special cases.
\subsubsection{Power-law expansion :}	
In power-law model the evolution of the scale factor of the universe yields $a(t)=a_0 t^D$, where $a_0$ and $D$ are constant parameter. In RS  regime with power-law expansion, the  expression of density of energy  and bulk viscous pressure yield, respectively
	\begin{equation}
	\rho=\rho_{0} D t^{-1}, \;\;\;\; \;\;\Pi = -\rho_{0} [(1+\omega)D-\frac{1}{3}] t^{-1} ,
	\end{equation}
	where $D>\frac{1}{3(1+\omega)}$ for physically relevant solutions ($\Pi<0$). 
 In RS regime  for power-law expansion with Full Israel Stewart (FIS)  theory, the field Eq. (28) yields
   \begin{equation}
	B_1 + B_2 t^{r} + B_3 t^{(r-s)} =0,
	\end{equation}             
where $B_1=\frac{9}{2}(1+\omega)D^2-\frac{3D}{2}((1+\omega)(r-s)+2)+\frac{1}{2}(r-s)+\frac{1}{2(1+\omega)}$, $B_2=\frac{1}{\beta}\rho_{0}^{1-r} D^{1-r} [3(1+\omega)D-1] $ and $B_3=-9\eta\rho_{0}^{s-r} D^{s-r+2}$. In RS regime power-law solutions are permitted in the following cases:\\
Case (i) $ r\neq 0$ and $s\neq r$: Power-law solution is permitted for $B_1=B_2=B_3=0$. It  shows $\rho_{0}=0$ which is not  physically accepted solution.\\
Case (ii) $ r\neq 0$ and $s=r$: In this case, power-law solution is permitted for $B_1+B_3=0$ and $B_2=0$. It  shows the power-law exponent $D=\frac{1}{3(1+\omega)}$. It indicates that the value of bulk viscous stress $(\Pi)$ become zero, which is also not physically acceptable power-law solution.\\
Case (iii) $ r\neq 0$ and $s=0$: In this case, power-law solutions are permitted for $B_1=0$ and $B_2+B_3=0$. It  shows the bulk viscus constant $\beta_1= \frac{(3(1+\omega)D-1)\rho_{0}}{9D}$ and the power-law exponent  $D=\frac{r}{3}+\frac{1}{3(1+\omega)}$ or $\frac{1}{3(1+\omega)}$. Here we note that  physically viable solution are permitted for $D=\frac{r}{3}+\frac{1}{3(1+\omega)}$. At this particular situation  the values power law exponent $(D)$ depends on  bulk viscous parameter $r$ and equation of state parameter $(\omega)$.  If  we choose $r=5+\frac{\omega}{1+\omega}$ as a special case, it yields power-law exponent $D=2$. Hence, the  deceleration parameter $(q)$ becomes  $q=-0.5$. In this context, One could note that the present value of deceleration parameter $(q)$ is very closed to $q\sim-0.5$. \\
Case (iv) $r=s=0$:  In this case,  power-law evolutions  are  acquired   in the RS dominated regime and  FIS theory for $B_1+B_2+B_3=0$ which leads to
\begin{equation}
9\left[1+\omega-2\eta+\frac{2(1+\omega)\rho_{0}}{3\beta}\right] D^2 -2\left[3+\frac{\rho_{0}}{\beta}\right]D +\frac{1}{1+\omega}=0.
\end{equation}
The power-law dominated inflation is permitted in Full Israel Stewart theory with known values of other parameters in Eq. (31).
 Figure 7 shows the variation of  $D$ vs.  $\omega$ for various $\beta$. Shadow section of Fig. 7 represents physically unsuitable ($\Pi>0$) region in power law type evolution. The figure  suggests that high values of $\omega$ and $\beta$ are appropriate for power law inflation in RS dominated regime.
\begin{figure}[ph]
\centerline{\includegraphics[width=2.0in]{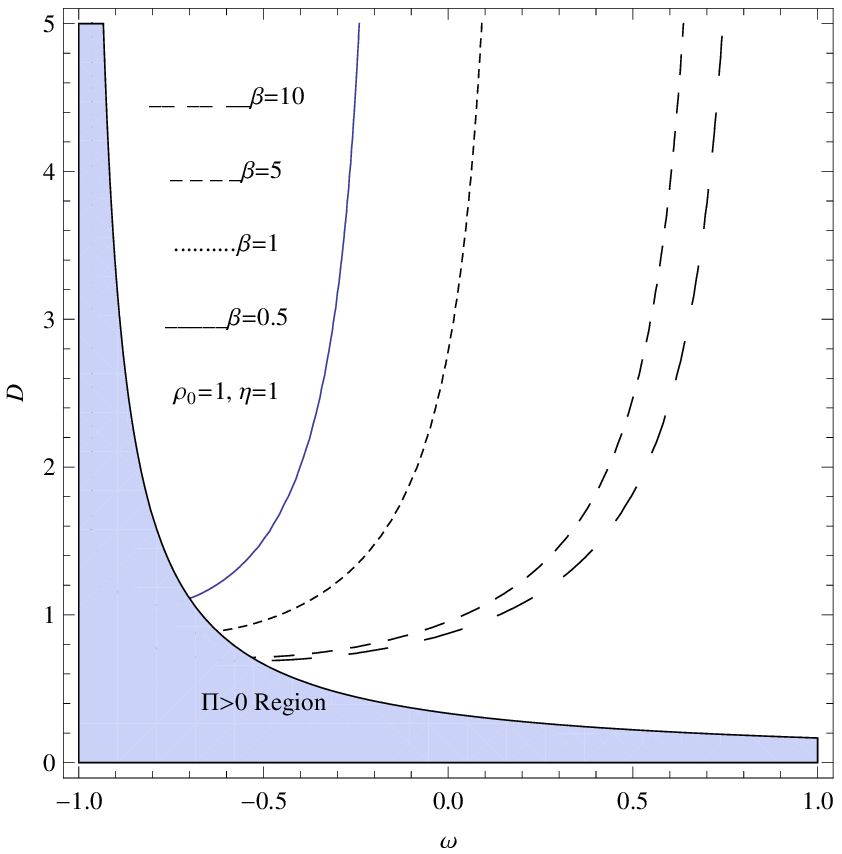}}
\vspace*{8pt}
\caption{Illustrates  $D$  vs  $\omega$ in RS dominated regime for various $\beta$ with other known parameters. \protect\label{fig7}}
\end{figure}
\\
 In the absence of bulk viscosity ($ \beta =\beta_1=0$) the field equation ($\dot{H}+3(1+\omega)H^2=0$) in RS regime 
		yields power law type expansion ($a(t)=a_0 t^D$) unless $\omega=-1$, where power law exponent $D=\frac{1}{3(1+\omega)}$. In RS regime without bulk viscosity, power law type accelerated expansion ($D>1$) is permitted for $-\frac{2}{3}>\omega>-1$. 
\subsubsection{Exponential expansion:}
Exponential type cosmic model $(a(t)\sim e^{H_1 t})$ is permitted in RS brane  for $\ddot{H}=\dot{H}=0$ or $H=const.=H_1$ of Eq. (28). Exponential expansion leads to de Sitter type expansion for $H_1>0$ even in the presence of matters.
In exponential expansion  model the Hubble's parameter in RS region  yields
\begin{equation}
\frac{(1+\omega)}{3\beta}\rho_{0}^{1-r}H_1^{-r} - \eta\rho_{0}^{s-r}H_1^{s-r} +\frac{(1+\omega)}{2}=0.
\end{equation}
In RS regime for exponential expansion,  following cases for simplicity are studied:\\
Case (i) $r=s$ and $s\neq 0$ : The exponent $(H_1)$ of exponential expansion  yields $H_1 =\left[\frac{3\beta(2\eta-1-\omega)}{2(1+\omega)\rho_0^{1-r}} \right]^{-\frac{1}{s}}$. The de Sitter type expansion is permitted for  $2\eta-1>\omega>-1$. The causality ($\omega\leq1$) of the solution implies  following  constraint on the parameter $\eta$, which is $0<\eta\leq1$.\\
Case (ii) $r\neq 0$ and $s= 0$ : In this  case, exponential  exponent $(H_1)$  yields $H_1 =\left[\frac{3(1+\omega)\rho_0^r}{2(3\eta-\frac{1+\omega}{\beta}\rho_0)} \right]^{-\frac{1}{r}}$. Here exponential expansion is permitted  for the following constraint among the parameters $1\geq\omega>-1$ and $\beta_1 > \frac{(1+\omega)\rho_0}{3\eta}$. \\
Case (iii) $r=0$ and $s\neq 0$ : In this particular case exponent $(H_1)$ of exponential expansion  reduces to $H_1 =\left[\frac{(1+\omega)}{3\eta\rho_0^{s}}\left(\frac{\rho_0}{\beta}+\frac{3}{2} \right) \right]^{\frac{1}{s}}$. In RS regime, de Sitter type expansion is permitted here for  $1\geq\omega>-1$, $\eta>0$ and $\beta>0$.\\
 In the absence of bulk viscosity ($ \beta =\beta_1=0$) the field equation ($\dot{H}+3(1+\omega)H^2=0$) in RS regime 
		yield exponential expansion ($a(t)=e^{H_1 t}$) for $\omega=-1$, which permit exponential accelerated expansion for $H_1>0$.\\
Adopting the method for stability analysis of GB dominated regime, it is also possible to learn stability for the evolution in the RS dominated regime with dissipative effect.  The   exponential evolution admits   accelerating phase for $H_1>0$. 
To revise stability of cosmic evolution due to equilibrium points or fixed points in RS dominated regime with causal theory, one can rewrite Eq. (28) in term of two autonomous first order differential equations which are
\[ \dot{H}=y ,\]
\[ \dot{y}= P(y,H)= -\frac{1}{2}\left[r-s-\frac{3+4\omega}{1+\omega}\right] \frac{y^2}{H}-\left[ \frac{3}{2}(1+\omega)(r-s)+3+\frac{\rho_0^{1-r}}{\beta}H^{-r} \right] H y    \]
\begin{equation}
-3\left[\frac{1+\omega}{\beta}\rho_0^{1-r} H^{3-3r}-3\eta \rho_0^{s-r} H^{s-r+3} +\frac{3}{2}(1+\omega)H^3  \right]  .
\end{equation}                
The phase points  $P(0,H_1)=0$  in phase space are  distinguished by  $\dot{y}=y=0$. Following expansion of Taylor's series and linear approximation \cite{smith}  of fixed points, Eq. (28) becomes
\begin{equation}
\dot{y}= e' H+ h' y \;,
\end{equation} 
where the constants $e'=-3(1+\omega)H_1^2\left[\frac{(-s)}{\beta}\rho_0^{1-r} H_1^{-s} + \frac{3}{2}(r-s) \right]$ and $h'=-H_1 \left[\frac{3}{2}(1+\omega)(r-s)+ 3+\frac{1}{\beta} \rho_0^{1-r} H_1^{-r} \right]$.
The characteristics of fixed points associate to de Sitter type evolution in RS dominated regime are exposed within Table 2.  Stable accelerated exponential  evolution is permitted for (i) $r\geq s$, $s< 0$, $2\eta-1\geq \omega > -1$ and  $0<\eta\leq 1$ (ii)  $r > s$, $s\leq 0$, $2\eta-1\geq \omega > -1$ and  $0<\eta\leq 1$. 
\begin{table}[ht]
\tbl{ Stability  of fixed points associated to de Sitter type expansion in RS dominated regime for causal solution. }
{\begin{tabular}{@{}cc@{}} \toprule
$Constraints\;of\; parameters$ & $Type \;of\;equilibrium \;points $   \\  \colrule
$r\geq s$, $s<0$, $2\eta-1\geq \omega > -1$, and  $0<\eta\leq 1$  & $stable\; attractor$ \\
$r > s$, $s\leq 0$, $2\eta-1\geq \omega > -1$ and  $0<\eta\leq 1$ & $stable\; attractor$ \\
$r < s$, $s\geq 0$, $ \omega > -1$ and  $H_1 > 0$  & $unstable\; saddle$ \\
$r=s$, $s=0$ and  $H_1 > 0$   & $centre$ \\
$r<s$, $s<0$, $\omega>-1$, $H_1> \left[\frac{3(s-r) \beta}{2\rho_0^{1-r}}\right]^{\frac{-1}{r}} $& $stable\; attractor$ \\\botrule
\end{tabular}}
\end{table}

\subsubsection{Evolution in the vicinity of the stationary solution : }
In Eq. (32)  $H=H_1$ implies inflationary expansion with a constant rate given by $H_1$. To examine the behaviour of the scale factor and the  Hubble parameter  in the vicinity of the stationary solution analytically, one can consider the perturbation as $H=H_1 +\chi$ and $\chi<<H_1$. Using the relation the analytic behaviour of scale factor in the vicinity of stationary solution yields $a(t)=a_0 e^{H_1t+\int\chi dt}$. By Setting $H=H_1+\chi$, with $|\chi<<H_1|$ and after linearization Eq.(28)  yields,
\begin{equation}
\ddot{\chi}+ \lambda_3\dot{\chi}+  \lambda_4 \chi =0,
\end{equation}
 where we consider the constants  $\lambda_3=(\frac{15+12\omega}{2}+ \frac{H_1^{-r}}{\beta\rho_{0}^{r-1}})H_1$, $\lambda_4= 3(\frac{(1+\omega)(-s) }{\beta \rho_{0}^{r-1}}  H_1^{-r}+\frac{3(3+4\omega)}{2}) H_1^2$ and $r-s-\frac{3+4\omega}{(1+\omega)}=0$. The solution of above Eq. (35) yields 
\begin{equation}
\chi(t)=\chi_3 e^{\delta'_+ t} +\chi_4 e^{\delta'_-t},
\end{equation}
where $\chi_3$ and $\chi_4$ are constants which depend on initial conditions. Here $\delta'_+$ and $\delta'_-$ are the roots satisfying following relation
\begin{equation}
\delta'_{\pm}=\frac{\lambda_3}{2}\left[-1\pm\sqrt{1-\frac{4\lambda_4}{\lambda_3^2}}\right].
\end{equation}
Several possibilities arise and  following cases are consider for simplicity:\\  
Case (i) weak damping $(\lambda_3 ^2<4\lambda_4)$ : The quantity within  the square root in Eq. (37) will be negative and the corresponding solution for $\chi$ yields  $\chi(t)=\chi'(0)e^{-\frac{\lambda_3 t}{2}} cos(\frac{1}{2}\sqrt{\lambda_3^2-4\lambda_4}\;t)$ , where $\chi'(0)$ is a constant. Hence Hubble parameter ($H$) exhibits an oscillatory damped behaviour of frequency $n=\frac{1}{2}\sqrt{\lambda_3^2-4\lambda_4}$ around the stationary solution $(H=H_1)$. The damped oscillatory solution yields inflation  with oscillation given by the term $\int \chi dt={e^{-\frac{\lambda_3 t}{2}}[b_1 cos(\frac{1}{2}\sqrt{\lambda_3^2-4\lambda_4}\;t+ b_2 sin(\frac{1}{2}\sqrt{\lambda_3^2-4\lambda_4}\;t]}$, where $b_1$ and $b_2$ are constants. This nearly stationary solution has the curious feature \cite{pavon}. Stable expansion of the universe with time around a equilibrium points $(H_1)$ are permitted for  $\lambda_3^2<4\lambda_4$ and $\lambda_3>0$.   Hence the  solution will be stable spiral for $s\leq 0$ and $\omega>-\frac{3}{4}$. In RS regime for $r=0$, one can obtain a damped oscillatory behaviour around $H_1$ for the constraint among the parameters $\frac{\rho_0}{\beta}<\frac{21+36\omega}{2} \pm 6 \sqrt{8\omega^2+10\omega+3}$.\\ 
Case (ii) strong damping $(\lambda_3 ^2 > 4\lambda_4)$ : The quantity within the square root in Eq. (37) is real and both roots in Eq. (36) will be real. Furthermore,  the quantity within the square bracket in Eq. (37) is negative for $\lambda_4>0$. Hence both  solutions shall be stable node for $s\leq 0$ and $\omega>-\frac{3}{4}$.  The   solution yields inflation  with  the term $\int \chi dt=b_1'e^{\delta'_+ t} + b_2'e^{\delta'_- t} $. 
\begin{figure}[ph]
\centerline{\includegraphics[width=2.0in]{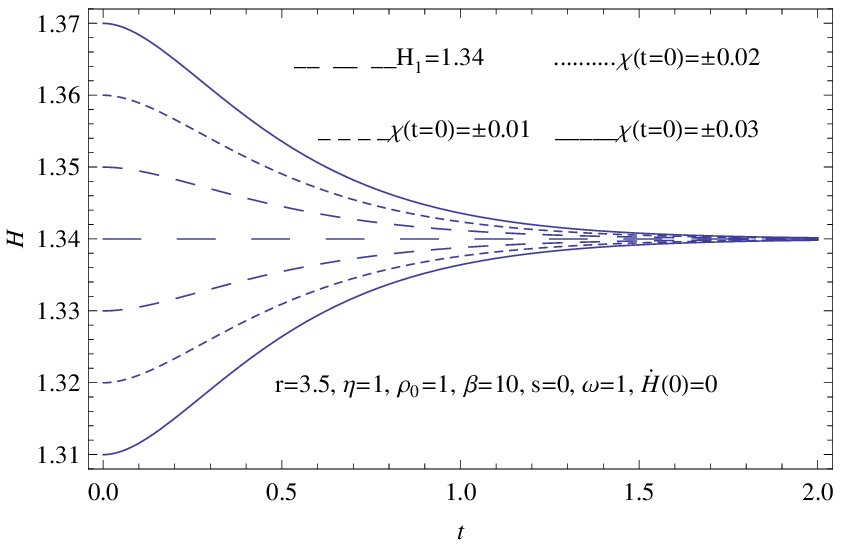}}
\vspace*{8pt}
\caption{Illustrates   $H$  vs.  $t$  in RS dominated regime for various  $\chi(t=0)$ for a known set  of parameters. \protect\label{fig8}}
\end{figure}
To examine stable expansion in the vicinity of the stationary solution, one can plot $H$ vs. $t$ for various  $\chi$ ($<<H_1$). Figure 8 shows $H$ vs. $t$ for a given value of other parameters in RS regime. The figure shows stable expansion in the vicinity of stationary solution $(H_1)$ for  $\lambda_3^2>4\lambda_4$ and $\lambda_4>0$.\\
Case (iii) critical damping  $(\lambda_3 ^2 = 4\lambda_4)$ : In this case $\delta'_+ =\delta'_-=-\frac{\lambda_3}{2}$, the solutions yield $\chi(t) = \left( b_1^{''} + b_2^{''} t \right) e^{-\frac{\lambda_3 t}{2}}$, here $b_1^{''}$, $b_2^{''}$ stand for arbitrary constants. The solutions resemble those for strong damping and the solutions show a  stable node  for  $s\leq 0$ and $\omega > -\frac{3}{4}$. \\
Case (iv) $\lambda_4<0$: In this case $\delta'_+$ and  $ \delta'_-$ are real but opposite sign. The solutions show a unstable saddle for (i) $s \geq 0$ and $-\frac{3}{4} > \omega > -1$, (ii) $s\leq 0$ and $\omega <-1$. 
\\
 In the absence of bulk viscosity ($ \beta =\beta_1=0$) the field equation in RS regime yields  $\dot{H}+3(1+\omega)H^2=0$ which is a first order differential equation of $H$. In the absence of bulk viscosity damped oscillatory behaviour of Hubble parameter ($H$)  in the vicinity of stationary solution ($H_1$) is not permitted  for the order of the field equation.
\section{Discussion}
In the article, we investigate  causal cosmological solutions for RS braneworld theory with  Gauss-Bonnet (GB) coupling. Here we consider the total effective cosmic pressure $(p_{eff})$ contains two part, namely, isotropic pressure ($p$) part along with bulk viscous pressure ($\Pi$) part. The isotropic part of fluid ($p$) is explained by a EoS which yields $p=\omega \rho$, where $\rho$ is   the energy density and $\omega$  is EoS parameter. The bulk viscous stress ($\Pi$) is explained by a causal theory, namely,  Full Israel Stewart (FIS). 
In the causal cosmology the  relaxation time and bulk viscous coefficient  are expressed  by respectively  $\tau=\beta \rho^{r-1}$ and $\zeta =\beta_1 \rho^s$, where $\beta (\geq 0),\;\beta_1(\geq 0),\; s(\geq0)$ and $ r(\geq0)$ are  constant parameters.
 The field equations that govern causal cosmological solutions in GB and RS regime are very nonlinear for find an universal analytical solution.
 For acquiring  analytic solutions of causal cosmology in GB and RS regime we consider following cases. Case (1)  $r=s=\frac{\epsilon -1}{\epsilon}$: The corresponding analytical solutions support cosmic exponential inflation in parametric form of time  for GB dominated and RS dominated regime, where $\epsilon=3$ and $\epsilon=1$ are in GB and RS regime respectively. Case (ii) $r=\frac{\epsilon -1}{\epsilon}$ and $s=\frac{\epsilon -3}{\epsilon},\;\frac{\epsilon -2}{\epsilon},\;\frac{\epsilon -1}{\epsilon}$: The corresponding analytic solution permits emergent universe model  both in GB and RS regime for some constraints among the parameters $(b_1, \;b_2,\;b_3,\;b_4)$. 
 We have also studied numerical solutions of cosmological parameters, namely, scale factor $a(t)$ and Hubble redshift parameter $H$ both for GB and RS regime.   Numerical solutions  in GB and RS regime are discussed respectively in Figs. 1-2 and Figs. 5-6 for  a given set of parameters. The Figs. 1 and 5 propose a declining nature of Hubble parametric ($H$) function with evolution ($t$). Figures 2 and 6 suggest that scale factor $a(t)$ is growing function among time $(t)$. 
 The figures show that in  GB regime the universe evolve more rapidly than RS regime  and for  a particular time, the values of scale factor are bigger with higher magnitudes of bulk viscous constant ($\beta$). 
 Analytic solutions of known from such as Power-law expansion, Exponential model and evolution in the vicinity of stationary solution are also discussed for GB dominated and RS dominated regime. Power-law solution $(a\sim t^D)$ is permitted with  $s=\frac{2}{3}$ and $s=0$ in GB and RS regime respectively. Figures (3) and (7) show plot of power law exponent ($D$) versus EoS parameter ($\omega$) with different  bulk viscous parameter ($\beta$).  The figures suggest that the opportunity for  Power-law acceleration increases with larger $\omega$ and  $\beta$ both for GB and RS regime. Exponential evolution  is permitted for GB dominated era and RS dominated era. The results of  stability  of fixed points with causal viscosity are summarized within Table 1 and Table 2 for GB and RS regime respectively.  Exponential  type stable  expansion is acquired for (i) $r\geq s$ , $s<\frac{2}{3}$ and $1\geq\omega>-1$ (ii) $r> s$ , $s\leq\frac{2}{3}$ and $1\geq\omega>-1$ in GB regime. For RS dominated era,  stable exponential expansion is permitted for $r< s$ , $s\leq 0$ and $1\geq\omega>-1$. 
We also analytically discuss cosmic evolution in the vicinity of the stationary solution. It is found that damped oscillatory behaviour of the Hubble parameter is permitted for causal    theory (FIS) both in GB and RS regime.  Figures 4 and 8 show  respectively $H$ versus $t$ for given other parameters with GB  dominated and RS dominated regime. The figures shows stable expansion of the universe with time around a stationary solution $(H_1)$ for different values of $\chi$ where $|\chi << H_1|$.  Stable evolution in the vicinity of the stationary solution is permitted for $s\leq\frac{2}{3},\; \omega>-\frac{1}{4}$,$\; r=s+\frac{1+4\omega}{3(1+\omega)}$ and $s\leq 0,\; \omega>-\frac{3}{4}$, $\; r=s +\frac{3+4\omega}{(1+\omega)}$ respectively in GB and RS regime.\\
However, in the absence of bulk viscosity ($ \Pi=0$), power law type accelerated expansion $(D>1)$ is permitted for $0>\omega>-1$ and $-\frac{2}{3}>\omega>-1$ in GB regime and RS regime respectively. The presence of bulk viscosity ($\Pi\neq 0$) may have several permitted range of $\omega$ for which power law type accelerated expansion is allowed for different values of $\tau$ and $\zeta$ in GB regime and RS regime. Again, in the absence of bulk viscosity exponential expansion is permitted only for $\omega=-1$ in GB regime and RS regime. While the presence of bulk viscosity may have several permitted values of $\omega$ for which exponential expansion is permitted for different values of $\tau$ and $\zeta$ in GB regime and RS regime.\\
In conclusion, it is shown that damped oscillatory behaviour of the Hubble parameter is permitted in the vicinity of the stationary solution for Full Israel Stewart (FIS) theory both  in GB and RS regime. It is also observed that stable stationary solutions are permitted in GB regime for (i) $r\geq s$ , $s<\frac{2}{3}$ and $1\geq\omega>-\frac{1}{4}$  (ii) $r> s$ , $s\leq\frac{2}{3}$ and $1\geq\omega>-\frac{1}{4}$  and that in RS regime for  $r< s$ , $s\leq 0$ and $1\geq\omega>-\frac{3}{4}$. Causal cosmology in RS brane including GB term 
allows Power-law type acceleration with higher  equation of state parameter ($\omega$) as well as bulk viscous constant $(\beta)$. We also note  down, the incorporation of Gauss Bonnet coupling in the Randall-Sundrum brane-world tends to enhance  the cosmic evolution.  
 In this context, we would like to mention that Power-law and Exponential models are permitted in GB and RS regime both in the presence and absence of viscosity. However, due to incorporation of causal viscosity in GB and RS regime, one can also obtain damped oscillatory behaviours of Hubble parameter in the vicinity of stationary solution.
\section*{Acknowledgement}
The author  acknowledge his gratitude to IUCAA, Pune and  IRC, NBU for widen the essential research amenities to begin the work.  He would also like to thank the anonymous reviewers for their important productive  remarks to improve the paper.

\end{document}